# Novel Physical Vapor Deposition Approach to Hybrid Perovskites: Growth of MAPbI$_3$ Thin Films by RF-Magnetron Sputtering

Sara Bonomi,[a] Daniela Marongiu,[b] Nicola Sestu,[b] Michele Saba,[b] Maddalena Patrini,[c] Giovanni Bongiovanni,[b] Lorenzo Malavasi [a,*]

[a]Department of Chemistry, University of Pavia and INSTM, Viale Taramelli 16 Pavia, 27100, Italy; [b]Department of Physics, University of Cagliari, S.P. Monserrato-Sestu km 0.7 Cagliari, 09042, Italy; [c]Department of Physics, University of Pavia and CNISM, Via Bassi 6 Pavia, 27100, Italy

**Corresponding Author**

Lorenzo Malavasi: lorenzo.malavasi@unipv.it

**ABSTRACT**

Solution-based methods represent the most widespread approach used to deposit hybrid organic-inorganic perovskite films for low-cost but efficient solar cells. However, solution-process techniques offer limited control over film morphology and crystallinity, and most importantly do not allow sequential film deposition to produce perovskite-perovskite heterostructures. Here the successful deposition of $CH_3NH_3PbI_3$ (MAPI) thin films by RF-magnetron sputtering is reported, an industry-tested method to grow large area devices with precisely controlled stoichiometry. MAPI films are grown starting from a single-target made of $CH_3NH_3I$ (MAI) and $PbI_2$. Films are single-phase, with a barely detectable content of unreacted $PbI_2$, full surface coverage and thickness ranging from less than 200 nm to more than 3 μm. Light absorption and emission properties of the deposited films are comparable to as-grown solution-processed MAPI films. The development of vapor-phase deposition methods is of interest to advance perovskite photovoltaic devices with the possibility of fabricating perovskite multijunction solar cells or multicolor bright light-emitting devices in the whole visible spectrum.

**Introduction**

Simple fabrication routes represent a major advantage of hybrid organic-inorganic perovskites for the manufacturing of efficient yet low-cost solar cells. Solution-based methods are the most widespread approach to prepare perovskite thin films, being at the same time reliable and cost-effective.[1-5] However, even optimized solution-process methods are affected by shortcomings. One is a lack of control over the low-temperature crystallization process, which is affected by many factors such as solvents and precursors, surface properties of the substrate, solvent evaporation during the deposition and annealing conditions, often leading to poor reproducibility of films morphology, thickness, crystallinity, and crystal size, properties that in turn have crucial influences on the photovoltaic performance. The second major shortcoming, more fundamental than the first one, is that sequential film deposition from solution cannot produce perovskite-perovskite heterostructures, since the solvent employed in depositing subsequent layers washes away the underlying ones. As a consequence, multijunction tandem solar cells and *p-n* junctions all perovskite based are advancing very slowly.

A very promising alternative to solution-based methods are the vapor-based deposition techniques, which started in the last few years to attract significant interest as a possible route to overcome the aforementioned problems.[6-9] In general, these methods are expected to provide purity of precursors and deposited films due to the vacuum environment and fine control of the deposition parameters, resulting in a high level of perovskite crystallization and reproducible films. Furthermore, vapor methods are suited for a scale-up preparation and large area deposition. It is crucial that vapor methods do not require the use of solvents and of annealing steps, allowing perovskite-on-perovskite deposition to create heterostructures and junctions.

To date, the vapor-based methods applied to the synthesis of hybrid perovskites are mainly based on vacuum evaporation process and vapor-assisted solution processes (VASP), with few other attempts of flash evaporation and ultrasonic spray coating.[7,8] We propose here a route based on a sputtering technique to provide highly reproducible single-phase hybrid perovskite films, full coverage of substrate surface, with the added bonus of being an industry-tested technique for large area film growth. The relatively lower deposition efficiency of sputtering is overcome by a magnetron-based device. To date, no reports about the deposition of hybrid perovskites thin films by sputtering has been reported.

Motivated by the above reported issues related to the preparation of hybrid organic-inorganic perovskites films, in this paper we demonstrate the successful one-pot growth of methylammonium lead iodide $CH_3NH_3PbI_3$ (MAPI) films by means of RF-magnetron sputtering starting from a single target made of a $CH_3NH_3I$ (MAI) and $PbI_2$ mixture with a 30% w/w excess of MAI. Sputtering technique allows to finely tune the deposition conditions by adjusting different parameters such as, for example, RF-power, gas pressure, and target to substrate distance thus providing an excellent platform to further optimize perovskite films as well as to extend the approach presented here for MAPI to any other material of interest in the field of Perovskite Solar Cells (PSCs).

**Results and Discussion**

Figure 1a presents a sketch of the MAPI thin films growth method used in the present work, *i.e.* the RF-magnetron sputtering starting from a target made of MAI and $PbI_2$ with a MAI excess of 30 wt%. Depositions were carried out at a RF-power of 40 W, with argon ($P=2\times10^{-2}$ mbar) as the sputtering gas in the DC-bias mode by setting its value to 80 V.

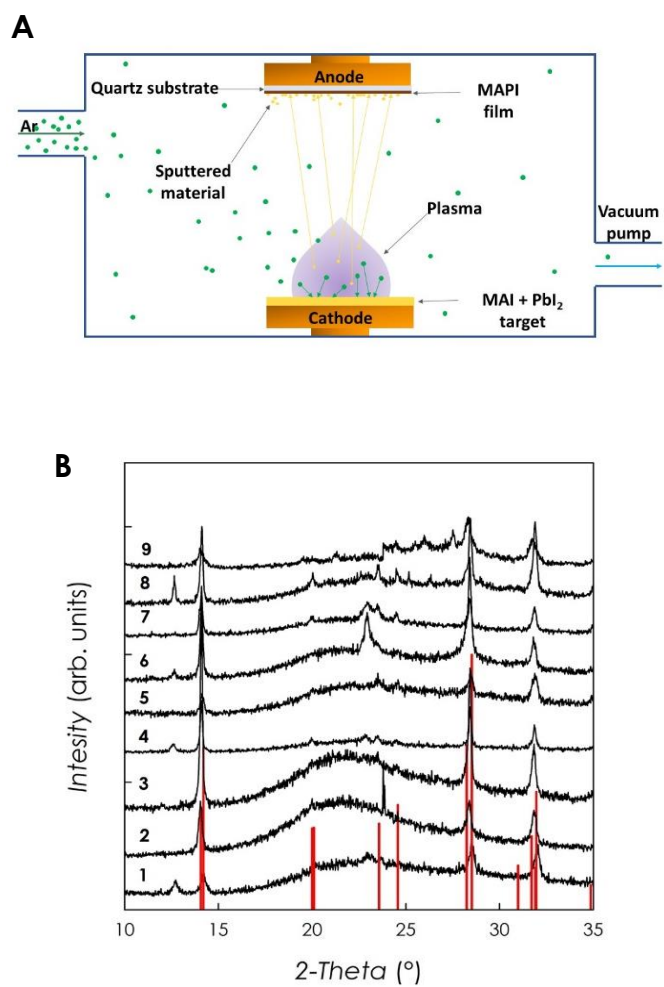

**Figure 1** – a) Schematic representation of the sputtering deposition method used in the present work to growth MAPI thin films; b) XRD pattern of MAPI films reported in Table 1.

Figure 1b shows the x-ray diffraction patterns (XRD) of a series of representative MAPI films with variable thicknesses from below 200 nm (Film 1) up to about 3.2 μm (Film 9) (details are reported in Table 1). Optimal growth conditions were obtained after extensive optimization work and are reported in the Experimental Section. The starting target for the sputtering deposition was a mixture of MAI and $PbI_2$ with a 30% w/w excess of MAI.

**Table 1**: Thicknesses and PL lifetimes of the MAPI films investigated in the present work

| Film | Thickness (nm) | Lifetime (ns) |
|------|----------------|---------------|
| 1    | < 200          | 1.2           |
| 2    | 220            | 1.9           |
| 3    | 300            | 1.7           |
| 4    | 350            | 4.3           |
| 5    | 440            | 13.2          |
| 6    | 510            | 2.7           |
| 7    | 640            | 5.6           |
| 8    | 940            | 1.8           |
| 9    | 3200           | 2.7           |

The films reported in Figure 1b have a crystal structure compatible with that of tetragonal MAPI (vertical red bars in Figure) and are single-phase or present very low $PbI_2$ impurities (below 5%). In addition, the diffraction peaks are quite narrow, indicating a good crystallization due to the sputtering process which is a significant result considering that the substrate is not heated during the deposition. The possibility of growing crystalline MAPI thin films on any substrate without in-situ and/or ex-situ thermal treatments, together with the use of a single target, are unique advantages of the present deposition process. In general, from Figure 1b, it can be observed a slight preferential growth along the (00*l*) directions, as suggested by the relative intensity of the experimental peaks corresponding to the (002) and (004) reflections compared to the calculated intensities. No significant shifts of the peaks as a function of film thickness are evident in the XRD patterns.

The surface morphology of the MAPI films has been investigated by Atomic Force Microscopy (AFM). Figure 2 reports some selected images of films with variable thickness.

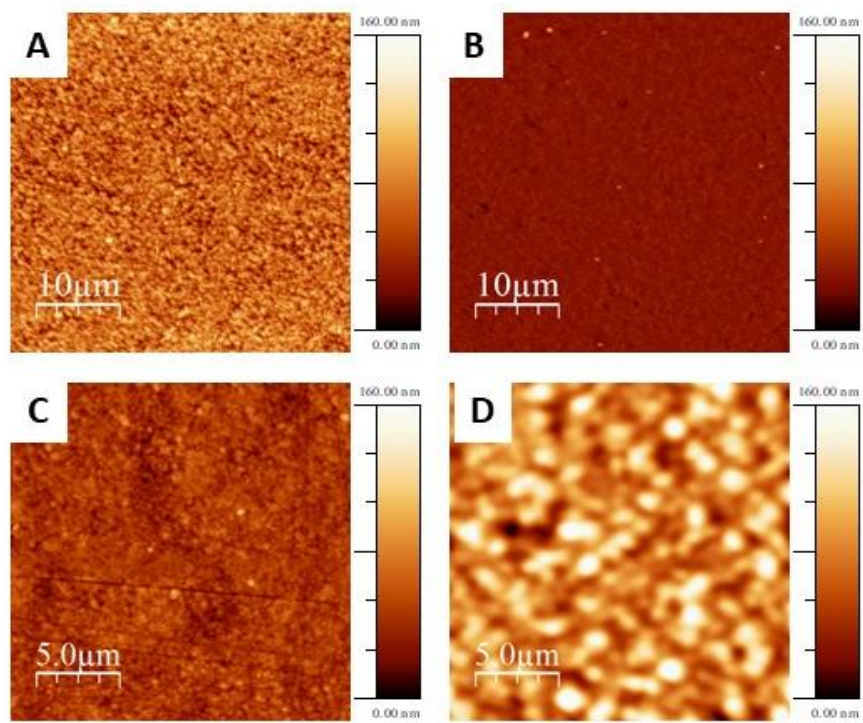

**Figure 2** AFM topography images of MAPI film over 40×40 μm for film **1** (A), **4** (B), **7** (C) and **9** (D).

From the images of Figure 2, it is possible to see complete coverage of the substrate surface starting from film 4 (Figure 2B) which has an estimated thickness of about 340 nm. The morphology of the deposited layers is made of small spherical grains with size in the 100-200 nm range which tends to increase by increasing the film thickness. In particular, figure 2D shows that, for a bulk-like film (thickness ~3200 nm), the grain size is comparable to that of a polycrystalline powder. The average roughness estimated from the AFM maps, and defined as root mean square ($R_{RMS}$) of surface height, is around 20 nm for film 1 (thickness <200 nm), 7 nm for film 4 (thickness ~340 nm), 10 nm for film 7 (thickness ~640 nm) and 20 nm for film 9 (thickness ~3200 nm), values low enough to suggest a deposition mechanism based on layer growth on the fused silica substrate used here.

Optical and excited-state properties of the deposited film have been assessed by absorptance (A) and photoluminescence (PL) measurements. Fig. 3 shows a representative absorptance spectrum of a thin MAPI film (4) grown by sputtering. The spectrum of thicker samples clearly showed saturation phenomena for increasing photon energies above the bandgap, caused by the strong increase of the band-to-band absorption coefficient (see Fig. 1, Supporting Information). As expected, a sharp absorption edge is found in the typical region of MAPI bandgap, *i.e.* around 1.6 eV.[10] No evidence of enhanced light absorption at 550 nm due to PbI$_2$ is present, confirming its absence or, when present, its relatively low amount according to the XRD analysis. The measured photoluminescence spectrum is also reported in Fig. 3. The spontaneous emission is almost resonant with absorption, suggesting an intrinsic origin. Small variations of the emission peak energy from sample to sample, in the range of 10-20 meV, were observed (see inset of Fig. 3); an analogous effect was indeed reported on solution-processed MAPI thin films, too.[11] The measured photoluminescence spectrum was compared with the expected spectrum according to the reciprocity relation between absorption and emission: $PL(\hbar\omega) \propto A(\hbar\omega)\,\omega^2 e^{-\frac{\hbar\omega}{k_B T}}$.[12] As the strong absorptance is due to direct band-to-band transitions, the photoluminescence spectrum derived in this way stems from spontaneous photon emission involving the very same intrinsic states. The excellent agreement between the expected and the measured spectra (red and black curves in Fig. 3) further supports the absence of any extrinsic contribution to light emission in MAPI films grown by sputtering.

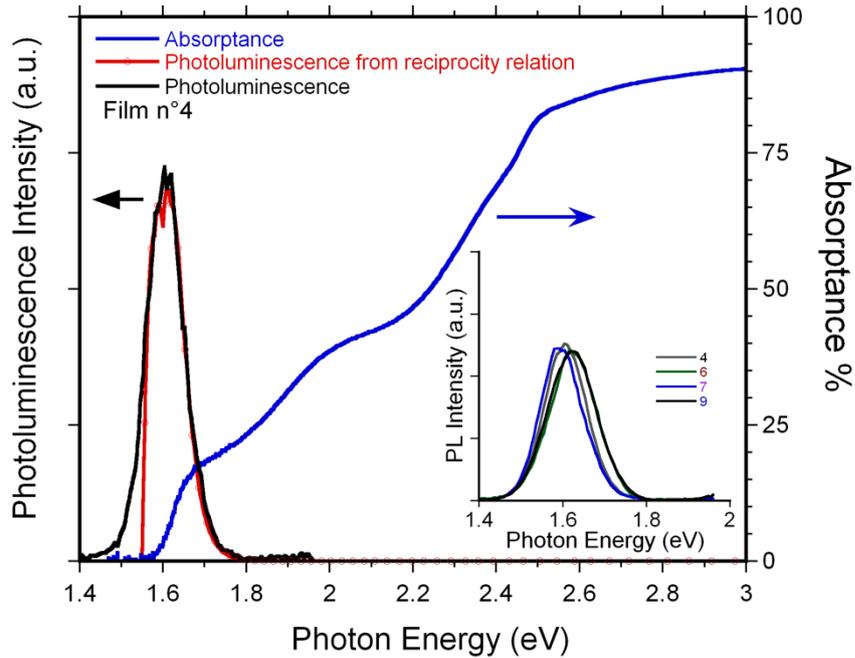

**Figure 3.** Photoluminescence and absorptance spectra of MAPI films. Main panel. Blue line: absorptance (A) spectrum of the film n°4. The black and red lines are the photoluminescence spectra. The former is the directly measured spontaneous emission spectrum. The second one is the emission spectrum calculated from the absorptance by using the reciprocity relation; the sharp drop of the calculated emission intensity at the low energy side of the spectrum stems from the experimental noise of the absorptance baseline. Inset. Directly measured photoluminescence spectra of various films.

Fig. 4 reports the dependence of the photoluminescence intensity on the excitation power level of two films (4 and 12). The reported behavior is representative of the whole set of samples investigated in this work: in all films, the spontaneous emission scaled linearly ($m=1$) with the laser intensity at low excitation regimes, then grew superlinearly at higher pumping intensities, following a $m=3/2$ power-law. In absence of nonradiative processes, light emission intensity is expected to linearly scale with the excitation level, independently of the nature of the involved electronic states, as all absorbed photons are subsequently reemitted. Nonradiative channels change this functional dependence. We have recently developed a simple approach to relate nonradiative processes with the dependence of the photoluminescence intensity on the excitation power.[13] In simple terms, if the concentration of electrons (holes) as a function of the laser intensity scales as a power law with index $m_e$ ($m_h$), the ensuing photoluminescence intensity also follows a power-law, but with index $m= m_e+ m_h$. We found that the kinetics triggered by deep traps leads to $m= m_e+ m_h=1+1/2=3/2$, under the assumption (Shockley-Read-Hall model) that traps in their ground state can be filled by only one type of carrier (in the following we assume to be electrons).[13] This result can be intuitively understood by looking at the recombination paths sketched in Fig. 4. Electrons undergo a conventional first order decay process; their density is thereby linearly dependent on the laser intensity in steady-state ($m_e=1$). Holes recombine with trapped electrons, following a true bimolecular process; the square of the hole density is thus proportional to the laser intensity, which leads to $m_h=1/2$. The hole lifetime is much longer than that of electrons, owing to the fact that the bimolecular recombination is a slow process at low carrier concentration. As a consequence, the densities of electrons and holes are unbalanced: the concentration of holes is much larger, and the semiconductor behaves as *p*-doped. $m=3/2$ is indeed the most common response observed in solution-processed MAPI films.[14-17]

In order to explain the linear behavior observed at low excitation regime, we first note that $m=1$ in our samples cannot be explained in terms of radiative recombinations. At low power levels, the measured photoluminescence quantum yield was in fact around $10^{-3}$, proving that nonradiative recombinations are the most efficient decays. Fig.4 shows the proposed model to explain the experimental findings. We assume that films contain a low concentration of shallow acceptors (unintentional chemical *p*-doping). As long as the excitation rate is low, the hole concentration is independent of excitation ($m_h=0$), while the dynamics of electrons remains unaffected ($m_e=1$) and thus $m=m_e$. Increasing the excitation level, the concentration of photogenerated holes becomes dominant and the $m=3/2$ behavior is recovered. This analysis is consistent with the observation that laser intensity, at which the transition from chemical doping to photodoping occurs, depends on sample, ranging from 10 mW/cm$^2$ (*i.e.*, below the light intensity delivered by sun) to a few $10^2$ W/cm$^2$. Unintentional doping (both *p*- and *n*-type) is often reported also in solution-processed MAPI films.[18,19]

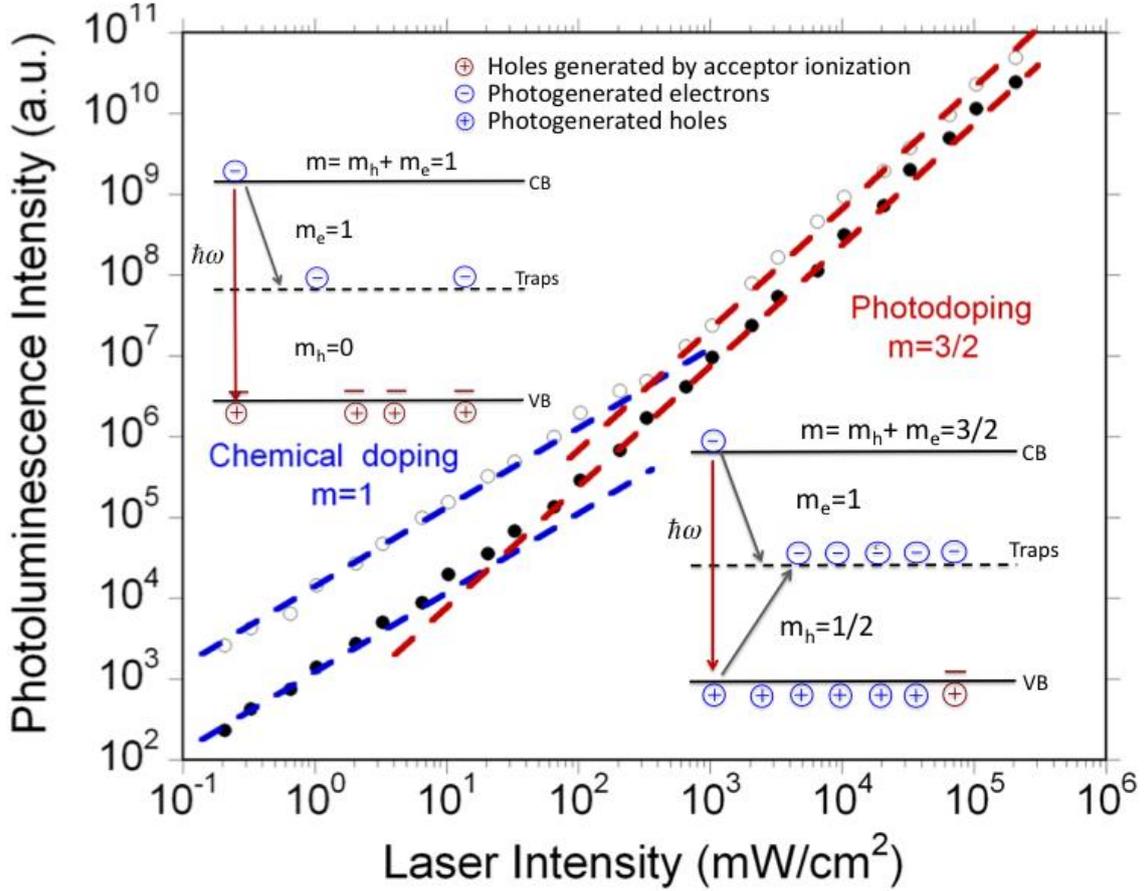

**Figure 4.** Photoluminescence dependence on excitation light intensity. Log-log plot of the photoluminescence intensity versus laser intensity. Full circles: film n°4; Empty circles: film n°12. The photoluminescence signal follows a power law as a function of the laser intensity. At low excitation density, the power index is m=1; at higher excitation densities, m=3/2. Insets: recombination processes at low and high intensities. Electrons and holes mostly decay non-radiatively via mid-gap energy traps (dashed line), which are assumed to capture only one type of carrier. At low excitations, the majority of holes are generated by ionization of shallow acceptors (unintentional chemical doping). At high excitations, the majority of holes are generated by light. According to the Shockely-Read-Hall model, a high concentration of free holes in the valence band (VB) is created, which turns out as a sort of photodoping because most of electrons are trapped and the subsequent concentration of free electrons in the conduction band (CB) remains low.

The time-resolved PL spectra of samples, representative of thin, intermediate and thick samples, are shown in Fig. 5, together with the spectrograms of the two films with the longest lifetimes. The spectrograms of all films show no spectral shift of the emission with time (see the spectra recorded at two different delays in the left panel of Fig. 5), as expected from intrinsic band-to-band emission following carrier thermalization at the band edges.

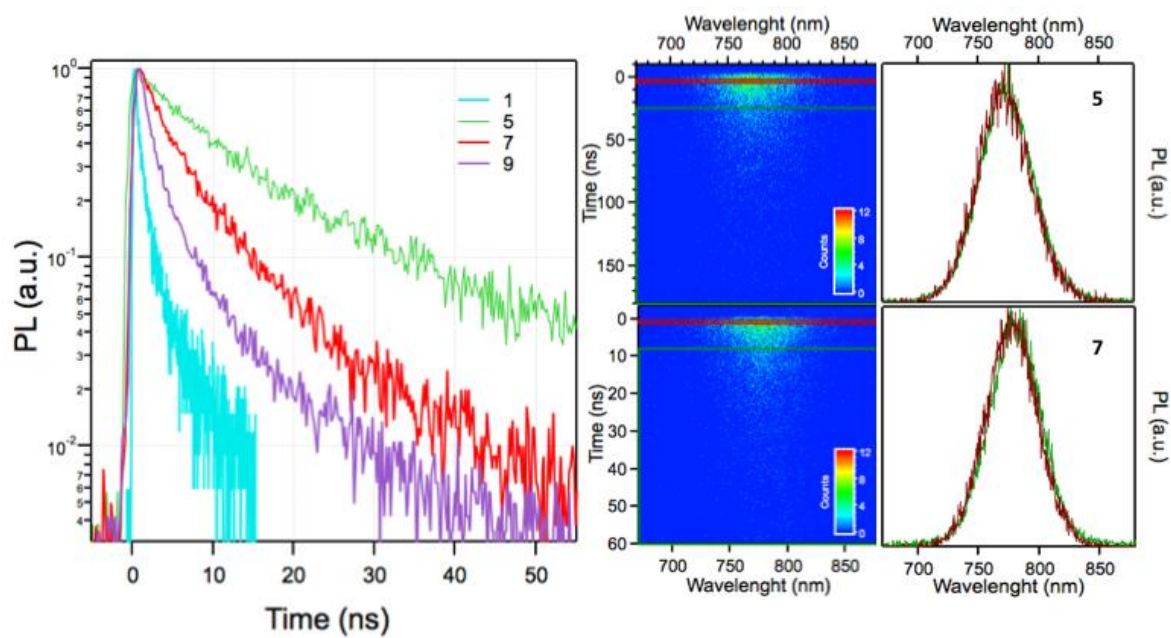

**Figure 5**. Time resolved photoluminescence. Left panel. Decay curves of the spectral-integrated photoluminescence signal for four films of different thicknesses. Central panels. Photoluminescence spectrograms of film 5 and 7, respectively. Right panels. Spectra of the spontaneous emission emitted in time window delimited by the red and green rectangles shown in the spectrograms reported in the central panels.

The photoluminescence lifetimes of all the films investigated in the present work are reported in Table 1 and provide a measure of trap concentration and non-radiative recombination rates associated to them. In general, the lifetimes are comparable to those of as-prepared solution-processed MAPI thin films,[20-24] but fall short with respect to state-of-the-art MAPI thin films optimized with post-growth treatments to passivate deep traps that have been demonstrated to enhance the photoluminescence lifetime and, consequently, quantum yield.[25-29] The results reported in this work refer to films which did not undergo any post-growth treatment and suggest that MAPI thin films produced by sputtering could be also substantially improved in their physico-chemical properties with future optimization work. By way of example, a possible morphology tuning and/or substrate heating to reduce the surface area of the film and/or to modulate the grain size, and therefore the surface defects, could already provide a significant increase in optical emission yield. Figure 5 shows that the longest carrier lifetimes have been obtained for intermediate thicknesses, which are close to the optimal values for PSCs absorbing layers. On other hand, for thinner and thicker films a reduction below 2 ns is observed. For thinner films this may be related to a major role played by surface defects, while for thicker film such reduction could be possibly related to an increase of the polycrystallinity which is known to enhance the nonradiative channels.[25-29]

## Conclusions

We reported the successful deposition of MAPI thin films by RF-magnetron sputtering. MAPI films were grown starting from a single-target made of MAI and $PbI_2$ (with a 30% w/w excess of MAI) and appeared to be single-phase, with full surface coverage and thickness ranging from less than 200 nm to more than 3 μm. The optical properties of the deposited films are comparable to as-grown solution-processed MAPI films and, in the future, the photoluminescence quantum yield could be substantially improved with post-growth passivation treatments. The development of vapor-phase deposition methods is of great interest in the current research on hybrid perovskites in view of a scale-up of device fabrication, the precise control of stoichiometry and the possibility of growing perovskite-perovskite heterostructures.

## Methods

*Film deposition:*

Thin films of MAPI have been deposited on amorphous silica substrates (MaTek, roughness ca. 1 nm) by means of radio frequency magnetron sputtering starting from a MAI/PbI$_2$ mixture (Aldrich, >99.9%) with a 30% w/w excess of MAI. Depositions parameters were: i) target-to-substrate distance, 2 cm, ii) RF-power, 40 W, iii) argon pressure, $2\times10^{-2}$ mbar. The depositions have been carried out in DC-bias mode by setting the value to 80 V. Film thickness has been determined by means of a mechanical profilometer. Estimated film growth is about 30 nm/min.

*XRD Diffraction:*

The structural properties of the deposited thin films were characterized by X-ray diffraction (XRD) by means of a Bruker D8 Advance instrument (Cu radiation) in a Bragg-Brentano set-up.

*Optical Properties Measurements:*

Reflectance (*R*) and transmittance (*T*) measurements were performed at 8° angle of incidence using a dual-beam spectrophotometer with an integrating sphere accessory (Agilent Cary 5000 UV-Vis-NIR). Absorptance (A) was calculated as 1-*R*-*T*.

Time-resolved photoluminescence. Samples were mounted excited by a regenerative amplifier laser (Coherent Libra) delivering 100-fs-long pulses at a repetition rate of 1 KHz. Photoluminescence was dispersed with a grating spectrometer (Princeton Instruments Acton SpectraPro 2300i equipped with a 50 gr/mm grating blazed at 600 nm), dispersed and detected by a streak camera (Hamamatsu)

CW photoluminescence. Samples were excited by a diode-pumped Nd:YVO$_4$ CW laser at 532 nm (Spectra Physics Millennia). Photoluminescence was dispersed by a grating spectrometer (Princeton Instruments Acton SpectraPro 2300i) and detected by a LN-cooled CCD camera (Princeton Instruments PIXIS). At the highest excitation intensities, laser beam was chopped to reduce the overall thermal loading.

*AFM Measurements*:

Surface topography and film roughness were measured by atomic force microscopy (AFM) with a NT-MDT Solver P47H-Pro instrument in semi contact mode at 1 Hz scan speed by a Etalon high-resolution non-contact silicon tip. Images were analyzed by WSxM software.

## Supplementary Information

UV-VIS transmittance spectrum of MAPI films.

## Competing interests

The authors declare no competing interests.

## Author Contribution

SB growth the films, DM performed the AFM and absorbance measurements, NS performed the PL measurements, MS and GB analyzed the optical measurements and wrote part of the paper; MP contribute with experimental design and LM devised the project, coordinated the work and wrote part of paper related to film growth and structural characterization.

## Acknowledgements

The authors gratefully acknowledge the project PERSEO-"PERovskite-based Solar cells: towards high Efficiency and lOng-term stability" (Bando PRIN 2015-Italian Ministry of University and Scientific Research (MIUR) Decreto Direttoriale 4 novembre 2015 n. 2488, project number 20155LECAJ) for funding.